\documentclass[useAMS,usenatbib]{mn2e}

\usepackage{graphicx}
\usepackage{amssymb}
\usepackage{color}

\newcommand{\nv}{}

\title[On the acceleration and deceleration of jets]%
{On the acceleration and deceleration of relativistic jets in active galactic nuclei -- II. Mass-loading}
\author[E.~E.~Nokhrina, V.~S.~Beskin]
{E.~E.~Nokhrina$^{1}$\thanks{E-mail: nokhrina@phystech.edu (EEN)} and V.~S.~Beskin$^{1,2}$
\\
$^{1}$Moscow Institute of Physics and Technology, Institutsky per.~9,
Dolgoprudny 141700, Russia\\
$^{2}$Lebedev Physical Institute, Russian Academy of Sciences,
Leninsky prospekt 53, Moscow 119991, Russia\\
}

\begin{document}

\date{Accepted \dots Received \dots; in original form \dots}

\pagerange{\pageref{firstpage}--\pageref{lastpage}} \pubyear{2017}

\maketitle

\label{firstpage}

\begin{abstract}
The effect of a mass-loading of the magnetohydrodynamic (MHD) flow in a relativistic jet from active galactic nuclei (AGN) 
due to $\gamma\gamma\rightarrow e^+e^-$ conversion is considered analytically. \nv{We argue that the effects of charge average separation
due to specific initial pairs' motion lead to partial magnetic and electric fields screening or enhancement. 
The effect of the fields screening has not been considered earlier.} The pairs with the center-of-mass moving faster 
or slower than the bulk plasma flow create a surface charge and a current which either screen or enhance both electric
and magnetic fields in a pair creation domain. This impacts the bulk flow motion, which either accelerates or decelerates.
The pairs with the center-of-mass moving with exactly the drift velocity does not induce the fields disturbance. In this case the flow
decelerates due to pure mass-loading. For these different cases the Lorentz factor of the loaded outflow is calculated as a function
of loading pairs number density. The effect may be important on sub-parsec to parsec scales due to conversion of TeV
jet radiation on soft infra-red to ultra-violet external isotropic photon field. This leads to an outer shell acceleration.
The conversion of MeV jet radiation on larger scales may account for the flow deceleration due to pure mass-loading.
\nv{The proposed mechanism may be a source of internal shocks and instabilities in the pair creation region.} 
\end{abstract}


\section{Introduction}
\label{s:intro}
 
The recent MOJAVE analysis of kinetic properties of the observed 
bright features of active galactic nuclei (AGN) has shown that jets have predominantly parallel 
to their velocity deceleration on scales 20 -- 100~pc of the order of $\dot{\Gamma}/\Gamma\sim 10^{-3}-10^{-2}$ per year
\citep{MOJAVE12},
which is hard to explain within the ideal magnetohydrodynamical (MHD) models. On the other hand, on scales
less than 10 -- 20 pc the bright features tend to accelerate at the same rate.  
In Paper I~\citep{BCh-16} 
we have proposed that an accurate account for the drag force may explain the observed plasma deceleration, the
characteristic length scales of deceleration being 100~pc. In this work we put forward another possible
mechanism for both jet acceleration and deceleration, i.e. the mass and charge loading due to two-photon pair creation.

The ideal MHD models of relativistic jets does not predict any radiation.
The cold plasma moves with exactly the drift velocity in crossed electric and magnetic fields 
\citep{Beskinbook, Tch09}, not undergoing any acceleration which may lead to radiation. Thus,
in order to introduce the radiation site, one needs to create some particle energy disturbance ---
either by internal shocks or by means of instabilities. The proposed in this paper effect
may account for such a site of radiation as well.  

This paper follows a work where the radiation drag is considered self-consistently as a mechanism for
plasma deceleration of a jet. Now we concentrate on charged pairs loading.
The mass loading itself has been considered by \citet{Lyu03} with the result of a flow acceleration
as a mass loading works effectively as a nozzle for the monopole magnetic field lines geometry. 
The result does not depend on a loaded particles charge
in contrast with the present paper. The work by \citet{DAKK03} proposed a mechanism that may account for extremely
efficient pairs acceleration by switching the neutral--charged state of particles. The series of papers
by \citet{SP06, SP08} explored also the so called photon breeding mechanism, in which $\gamma\gamma$
pair conversion is followed by efficient acceleration of pairs and consequent high energy radiation of 
photons (from where originates `breeding'). These works are very close to what is done here, since 
all of them also takes advantage of charged particle
behavior in electromagnetic field of a jet.

Thus, in this paper we consider the effects of charged pairs loading on the relativistic jet dynamics. For the 
two-photon $\gamma\gamma\rightarrow e^+e^-$ pair conversion, the opposite charges move in electric and magnetic 
field with relativistic velocities in contrast with bulk motion of cold plasma. If such pairs are
created in some jet domain (e.g., an outer shell), the local electric and magnetic field screening takes place due to
average charges separation and, consequently, appearance of electric and magnetic field due to these charges and the
corresponding currents. Locally, the drift velocity of plasma changes following the fields screening, and this leads either
to plasma acceleration due to MHD process, or plasma deceleration due to radiation of `excess' energy.

In Section~\ref{s:PL} we consider pairs created at rest in laboratory frame (i.e., in the reference
frame connected with 'central engine'), thus, moving bachwards with respect
to the jet bulk plasma outflow. We present the solution for such a 
case as being the most demonstrative and intuitively clear. For such simple initial conditions we
obtain the equation for the bulk Lorentz factor and the behavior of flow magnetization after loading.  
Starting from microscopic motion, we calculate the energy-momentum tensor of loading pairs in Subsection~\ref{ss:PM}. 
It is shown in Subsection~\ref{ss:MLFP} that the local screening of electric and magnetic fields is important 
and leads to bulk motion local deceleration and to the change in the integrals of motion as well. 
We estimate the appropriate change in a local magnetization of a flow in Subsection~\ref{ss:LFM}.
In Section~\ref{s:ArbVel} the result of general initial conditions for pairs is obtained.
We show that the pairs created in center-of-mass (c.m.) frame moving faster than a jet accelerate the bulk plasma
due to electric and magnetic field enhancement. If the pairs
are created with c.m. at rest in the jet proper frame, no field screening have place, so the flow
decelerates due to the pure mass-loading. The luminosities in high-energy jet radiation are estimated to
explain the observed by \citet{MOJAVE12} acceleration and deceleration rates.

\section{Set up}
\label{s:SU}

In what follows we will need a standard description of ideal axisymmetric MHD outflow. The \nv{magnetic field $B$
and electric field $E$} configuration is described by the magnetic flux function $\Psi(r,\;\varphi,\;z)$ in cylindrical
coordinate system \nv{with unit vectors $\left\{{\bf e}_{r},\,{\bf e}_{\varphi},\,{\bf e}_{z}\right\}$} by
\begin{equation}
\begin{array}{l}
\displaystyle{\bf B}=\frac{\nabla\Psi\times{\bf e}_{\varphi}}{2\pi r}-\frac{2I}{rc}{\bf e}_{\varphi},\\ \ \\
\displaystyle{\bf E}=-\frac{\Omega_{\rm F}}{2\pi c}\nabla\Psi,\label{defEB}
\end{array}
\end{equation}    
where $\Omega_{\rm F}(\Psi)$ \citep{Ferraro} is an angular velocity and one of five integrals, i.e. functions preserved
on the magnetic surfaces $\Psi={\rm const}$ (see, e.g., \citet{Beskinbook} for more detail). \nv{Here $c$ is the velocity of light.} The other four
are the entropy $s(\Psi)$, particle to magnetic flux ratio $\eta(\Psi)$, defined by
\begin{equation}
{\bf u}=\frac{\eta}{n}{\bf B}+\Gamma\frac{\Omega_{\rm F}r}{c}{\bf e}_{\varphi},
\end{equation} 
and energy and momentum fluxes
\begin{equation}
\begin{array}{l}
\displaystyle E(\Psi)=\frac{\Omega_{\rm F}I}{2\pi c}+\mu\eta\Gamma,\\ \ \\
\displaystyle L(\Psi)=\frac{I}{2\pi}+\mu\eta r u_{\varphi}.\label{defEL}
\end{array}
\end{equation}
Here the following physical properties of a flow are introduced: particle number
density in a jet proper frame $n$ \citep{Beskin-10}, the total current $I$ inside the magnetic tube, 
the flow bulk Lorenz factor $\Gamma$,
and the relativistic enthalpy $\mu$. In what follows we consider a cold flow $s=0$, so $\mu=m_{\rm e}c^2$.

The important function that characterizes an MHD outflow is a magnetization is the ratio
of Poynting flux to kinetic energy flux
\begin{equation}
\sigma=\frac{B^2}{4\pi m_{\rm e}c^2 n_{\rm lab}\Gamma},\label{sigma2}
\end{equation}
where $n_{\rm lab}$ is particle number density in the laboratory frame, and it can be expressed through the 
particle number density in a jet proper frame $n$ as $n_{\rm lab}=n\Gamma$.
Using the definitions (\ref{defEB}), the magnetization can can be rewritten  as
\begin{equation}
\sigma=\frac{\Omega_{\rm F}I}{2\pi m_{\rm e}c^3\eta\Gamma}\label{sigma3}.
\end{equation}

\section{Particle Loading --- Pairs Created at Rest}
\label{s:PL}

\subsection{Particle motion}
\label{ss:PM}

As was already stressed in the Introduction, the action of the pair loading on
the magnetically dominated flow was mainly considered either phenomenologically
\citep{DAKK03} or numerically \citep{SP06}.
Consistent analytical analysis for 1D (spherically symmetric) outflow was done
by \citet{Lyu03}. In particular, it was demonstrated that the action of particle
loading is similar to negative pressure. On the other hand, as will be shown below,
some important properties (such as anisotropic pressure and redistribution of
charges diminishing the electric and magnetic fields) were not included into consideration.
In the frame of MHD approached briefly outlined in Section~\ref{s:SU} this means that the
angular velocity $\Omega_{\rm F}$ and the total current $I$ change due to charge loading.  

Following \citet{Lyu03}, we consider here the simplest model of particle loading
when the electron-positron pairs are created at rest in the nucleus rest frame (laboratory frame).
This implies that these particle carry no energy $E$ and angular momentum $L$ flux, with total
gain in particle energy resulting from the diminishing of the Poynting flux of the magnetized
wind.

Let us solve the equation of motion
\begin{equation}
\frac{{\rm d}u^{i}}{{\rm d}\tau}=\frac{e}{m_{\rm e}c}F^{ik}u_k,
\end{equation}
for four-velocity of particle $u^i$ with the initial condition in nucleus rest frame
$u^i(0)=\left\{1,\;0,\;0,\;0\right\}.$
The solution is
\begin{eqnarray}
\gamma(\tau)&=&\frac{B^2}{B'^2}-\frac{E^2}{B'^2}\cos\Omega'\tau, \\
u^r(\tau)&=&-\frac{E}{B'}\sin\Omega'\tau, \label{Ur} \\
u^{\varphi}(\tau)&=&\frac{B_{\rm p}E}{B'^2}\left(1-\cos\Omega'\tau\right), \\
u^z(\tau)&=&\frac{B_{\varphi}E}{B'^2}\left(1-\cos\Omega'\tau\right).
\end{eqnarray}
Here $B$ and $E$ are the magnetic and electric fields in laboratory
frame, $B'=\sqrt{B^2-E^2}$ is a field in a jet frame, $\tau$ is the proper time in the 
charges comoving frame (which is different from the jet proper frame), and
\begin{equation}
\Omega'=\frac{eB'}{m_{\rm e}c}
\end{equation}
is a proper gyrofrequency, i.e. gyrofrequency in a jet proper frame instantly coincident with 
plasma bulk motion, which is essentially a drift motion in crossed magnetic and electric fields.

Assuming the pairs are created uniformly over the jet, we average the velocities and energy
in a given space point over all the particles with trajectories crossing this point. This is 
equivalent to averaging over the gyroperiod, with the procedure details described in  
Appendix~A. So, we obtain the following Lorentz factor and velocities:
\begin{equation}
\langle\gamma\rangle=\Gamma^2\left(1+\frac{\beta^4}{2}\right),
\end{equation}
\begin{equation}
\langle v_r\rangle=0,
\end{equation}
\begin{equation}
\langle v_z\rangle=c\beta\cos\alpha,
\end{equation}
\begin{equation}
\langle v_{\varphi}\rangle=c\beta\sin\alpha.
\end{equation}
Here we use the relation $B^2/B'^2=\Gamma^2$ with $\Gamma$ being the Lorentz factor of the drift velocity,
$\beta=\sqrt{1-\Gamma^{-2}}$ is a drift velocity, $\cos\alpha=B_{\varphi}/B$, and 
$\sin\alpha=B_{\rm p}/B$.

We stress that the mean energy of the individual loaded particle $\epsilon_{\rm ld}$ exceeds
essentially the energy
of particles in a jet $\epsilon_{\rm jet}=m_{\rm e}c^2\Gamma$:
\begin{equation}
\epsilon_{\rm ld}=m_{\rm e}c^2\langle\gamma\rangle\approx \frac{3}{2}m_{\rm e}c^2\Gamma^2.\label{e_loaded}
\end{equation}
However, the mean velocity of each loaded particle coincides with the drift velocity of plasma in a jet:
\begin{equation}
\langle{\bf v}_{\rm ld}\rangle={\bf v}_{\rm drift}.
\end{equation}
This implies that loading supplies the initially cold plasma with relativistic particles.

Let us determine the thermodynamical parameters calculating the components of energy momentum 
tensor
$T^{ik}_{\rm ld}$ of loaded particles. In order to do it, we average the components of the tensor
\begin{equation}
T^{ik}_{\rm ld}=m_{\rm e}c^2\langle n^{\rm rest}u^i u^k\rangle
\end{equation}
(see Appendix~B for more detail) 
and compare the result with the form of energy-momentum tensor expressed through the internal energy density
$\varepsilon_{\rm ld}$, longitudinal $P_n$ and transverse $P_s$ components of the pressure with respect
to the magnetic field \citep{Ts-95, Kuz-05}, and hydrodynamical four-velocity $U^i$:
\begin{eqnarray}
T^{ik}_{\rm ld}=\left(\varepsilon_{\rm ld}+P_{s}+\frac{{\bf b}^2}{4\pi}\right)U^{i}U^{k}
+\left(P_{s}+\frac{{\bf b}^2}{8\pi}\right)g^{ik}-
\nonumber \\
-\left(\frac{P_{s}-P_n}{{\bf b}^2}+\frac{1}{4\pi}\right)b^{i}b^{k}.
\label{Tik}
\end{eqnarray}
Here ${\bf b}^2$ is the plasma proper magnetic field energy density, and 
\begin{equation}
b^i=\frac{1}{2}\eta^{ijkl}U_jF_{kl},
\end{equation}
is the Lichnerowicz~\citep{Lich-67, AB-83} four-vector ($b^2=B'^2$).
It gives for the internal energy density of loaded particles
\begin{equation}
\varepsilon_{\rm ld}=m_{\rm e}c^2n_{\rm ld}\Gamma
\end{equation}
and for pressure components
\begin{equation}
P_n=0,
\end{equation}
\begin{equation}
P_s=\frac{1}{2}m_{\rm e}c^2n_{\rm ld}\beta^2\Gamma.\label{Ps}
\end{equation}
Hence, the energy-momentum tensor (\ref{Tik})
corresponds to anisotropic pressure with $P_{n} = 0$ resulting from particle
rotation in the $rz$-plane only. Thus, in this point
our approach differs from one considered by \citet{Lyu03}. It is important
that in the cylindrical geometry under consideration the volume force
${\cal F}_{i} = - \nabla_{k} P_{ik}$ appears even for constant value $P_{s}$:
\begin{equation}
{\cal F} = - \frac{P_{s}}{r}{\bf e}_{r}.
\end{equation}

\subsection{Mass-loaded flow properties}
\label{ss:MLFP}

The important question of a problem of pair loading the MHD flow is its impact on the 
angular velocity $\Omega_{\rm F}$, which describes the electric field, and the current $I$, which 
describes the toroidal component of the magnetic field. We argue that both quantities change due to 
charge loading, and its effect on the jet dynamics is significant and can not be reduced to pure redistribution of 
energy between the cold plasma with unchanged $\Omega_{\rm F}$ and $I$ and loaded plasma. 

\begin{figure}
\includegraphics[scale=0.4]{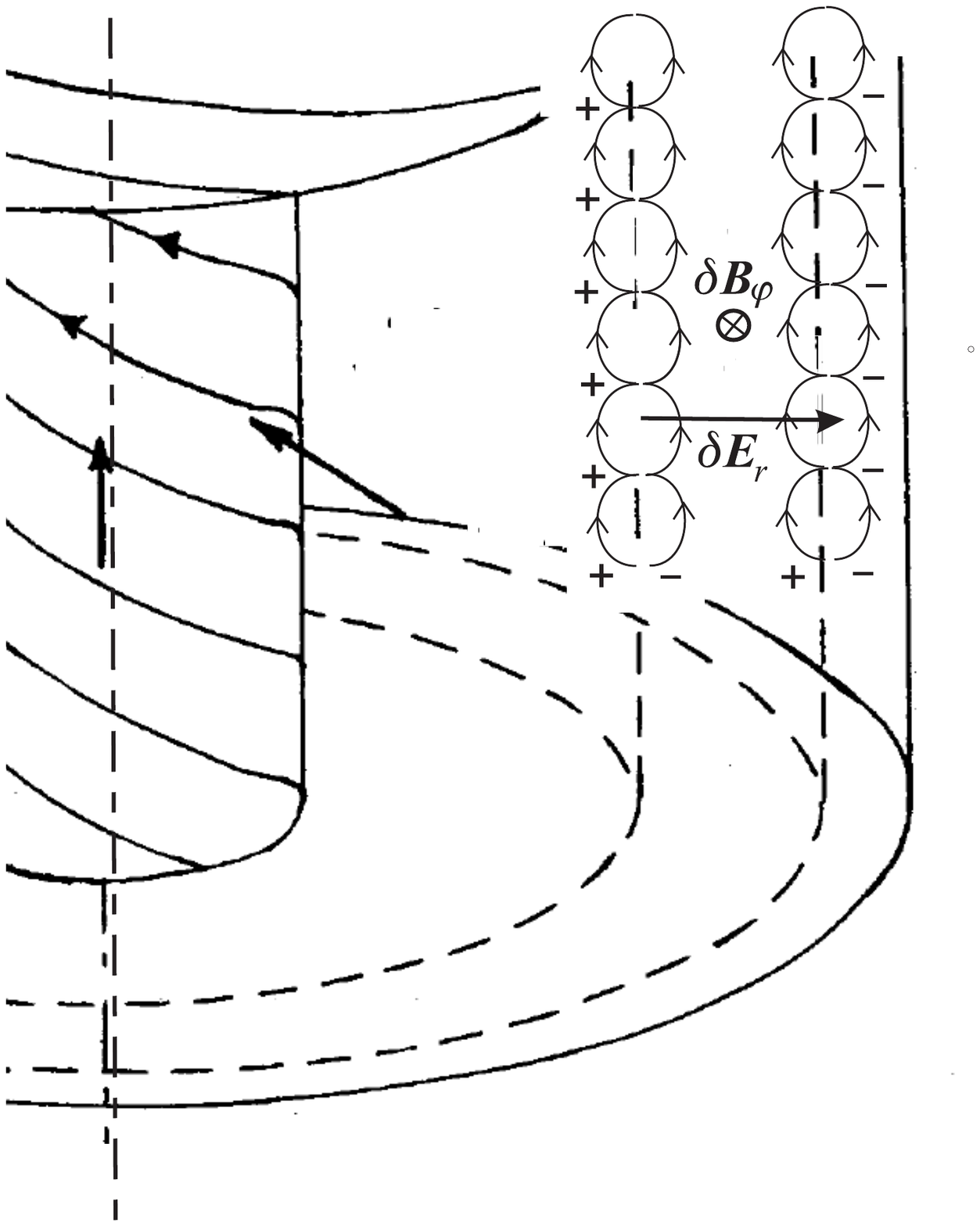} 
\caption{Cartoon of the loading plasma trajectories and induced electric and magnetic fields.
} 
\label{fig_1}
\end{figure}

Indeed, pairs initially created at rest start 
cycloid movement in crossed electric and magnetic fields with radial velocity component 
directed oppositely for electrons and positrons (it is obvious when one notices that the charge sign appears in gyrofrequency $\Omega'$). 
This leads to mean over gyroperiod separation (See Fig.~\ref{fig_1}) of positive and negative charges in $r$ direction
\begin{equation}
r_{\perp}=\langle r^+(\tau)-r^-(\tau)\rangle=-\frac{m_{\rm e}c^2{\beta}{\Gamma}^2}{e{B}}\left(2+{\beta}^2\right).\label{rperp0}
\end{equation}
In the simplest case
of a shell domain with uniform loaded particle number density $n_{\rm ld}$ as measured in a jet proper frame, the sides of constant $r$ possess   
a surface charge $\sigma_{\rm e}$ and a surface current
$\sigma_{\rm e}v_{\rm drift}$. 
Indeed, if pairs are created uniformly over a shell $\{L_{\rm r},\;r\Delta{\varphi},\;L_{\rm z}\}$ with mean separation $r_{\perp}$
in $r$ direction, this can be envisioned as two shells with uniformly distributed positive and
negative charges, shifted in $r$ direction on mean charges separation $r_{\perp}$. The total charges on 
opposite faces are $\pm enr_{\perp}r\Delta{\varphi}L_{\rm z}$, with effective surface charge density being
$\sigma_{\rm e}=enr_{\perp}$. As pairs are moving in crossed electric and magnetic fields with drift velocity,
there is a surface currents of opposite signs flowing along the $r$ boundaries of pair creation domain. The significant
magnetic field due to these currents is inside this domain. 
These charge and current densities screen partially the initial electric and magnetic fields. This 
must lead to a change in bulk jet velocity, depending on proportion in which fields were screened.

The effect of jet acceleration or deceleration in this model may be obtained self-consistently. We designate by
tilde all the flow characteristics after mass loading. They are the screened electric and magnetic fields, the appropriate Lorentz 
factor of a jet bulk motion and so forth. We assume that the proper motion of loaded particles is determined by 
these screened fields. \nv{In particular, we substitute into the expression for the 
average charge separation (\ref{rperp0}) the parameters of accelerated flow and the screened fields. 
Thus, in equation (\ref{rperp0}) we use the fluid velocity $c\tilde{\beta}$, Lorentz factor $\tilde{\Gamma}$, and
screened magnetic field $\tilde{B}$ instead of initial unperturbed flow parameters.} 

The surface charge in a laboratory frame is equal to
\begin{equation}
\sigma_{\rm e}=en_{\rm ld}^{\rm lab}r_{\perp}=e\tilde{\Gamma}n_{\rm ld}r_{\perp},
\end{equation}
where $r_{\perp}$ is a mean over laboratory time (or, equally, over the loaded particles ensemble) charges separation. 
The associated electric field disturbance for the given 
loaded particle number density is
\begin{equation}
\delta E=\frac{4\pi m_{\rm e}c^2n_{\rm ld}\tilde{\beta}^2\tilde{\Gamma}^3}{\tilde{E}}\left(2+\tilde{\beta}^2\right).\label{deltaElectric}
\end{equation} 
The appropriate magnetic field disturbance $\delta B$ produced by two opposite surface currents flowing along the 
domain boundaries of constant $r$ is $\delta B=\tilde{\beta}\delta E$, which gives 
\begin{equation}
\delta B=\frac{4\pi m_{\rm e}c^2n_{\rm ld}\tilde{\beta}^2\tilde{\Gamma}^3}{\tilde{B}}\left(2+\tilde{\beta}^2\right).\label{deltaMagnetic}
\end{equation}
The fields are disturbed only locally in the mass loading domain. The proposed self-similar way of calculating the
electric and magnetic fields is applicable while the disturbance is much less that the initial fields. So, we imply
that $\delta B/B\ll 1$, or equally 
\begin{equation}
\frac{1}{\sigma}\frac{n_{\rm ld}}{n}\Gamma\beta^2\left(1+\frac{\beta^2}{3}\right)\ll 1.\label{DeltaBB}
\end{equation}

Here let us note that such a screening will indeed lead to a jet deceleration. The initial fields 
correlate as $E=\beta B$. The fields created by loaded particles correlate as 
$\delta B=\tilde{\beta}\delta E$, so the magnetic filed screening is
less than that of electric. This leads to smaller drift velocity 
$\tilde{\beta}=(E-\delta E)/(B-\delta B)$.
 
Let us obtain the local (in pair creation domain) self-consistent drift velocity 
$\tilde\beta$ of bulk plasma due to fields screening. 
The fields depend now on $\tilde{\Gamma}$ and $n_{\rm ld}$:
\begin{equation}
\tilde{B}=\frac{B}{2}\left[1+\sqrt{1-\frac{4}{\sigma}\frac{n_{\rm ld}}{n}\frac{\tilde{\Gamma}^3}{\Gamma^2}\tilde{\beta}^2\left(2+\tilde{\beta}^2\right)}\right],
\label{tildeE}
\end{equation}
\begin{equation}
\tilde{E}=\frac{E}{2}\left[1+\sqrt{1-\frac{4}{\sigma}\frac{n_{\rm ld}}{n}\frac{\tilde{\Gamma}^3}{\Gamma^2}\frac{\tilde{\beta}^2}{\beta^2}
\left(2+\tilde{\beta}^2\right)}\right].
\label{tildeB}
\end{equation}
Here we use the expression for a flow magnetization given by (\ref{sigma2}).


The explicit equation on $\tilde{\Gamma}$ is
\begin{equation}
\tilde{\Gamma}=\frac{\tilde{B}}{\sqrt{\tilde{B}^2-\tilde{E}^2}},
\end{equation}
where $\tilde{B}=B-\delta B$ and $\tilde{E}=E-\delta E$ depend themselves through 
(\ref{tildeE})--(\ref{tildeB}) 
on $\tilde{\Gamma}$ and $n_{\rm ld}$.
After tiresome algebra one can obtain the following algebraic equation on $\tilde{\Gamma}$:
\begin{equation}
\begin{array}{l}
\displaystyle-16\tilde{\Gamma}^{10}-24\tilde{\Gamma}^9q\left(2\Gamma^2-1\right)+
\tilde{\Gamma}^8\left[32-9q^2\right]+\\ \ \\
\displaystyle+56\tilde{\Gamma}^7q\left(2\Gamma^2-1\right)+\tilde{\Gamma}^6\left[16\left(\Gamma^4-\Gamma^2-1\right)+24q^2\right]-\\ \ \\
\displaystyle-40\tilde{\Gamma}^5q\left(2\Gamma^2-1\right)+\tilde{\Gamma}^4\left[-16\left(\Gamma^4-\Gamma^2\right)-22q^2\right]+\\ \ \\
\displaystyle+8\tilde{\Gamma}^3q\left(2\Gamma^2-1\right)+4\tilde{\Gamma}^2q^2-q^2=0,
\end{array}\label{eq_for_gamma}
\end{equation} 
where 
\begin{equation}
q=\frac{4}{\sigma}\frac{n_{\rm ld}}{n}.
\end{equation}

\begin{figure}
\includegraphics[scale=0.4]{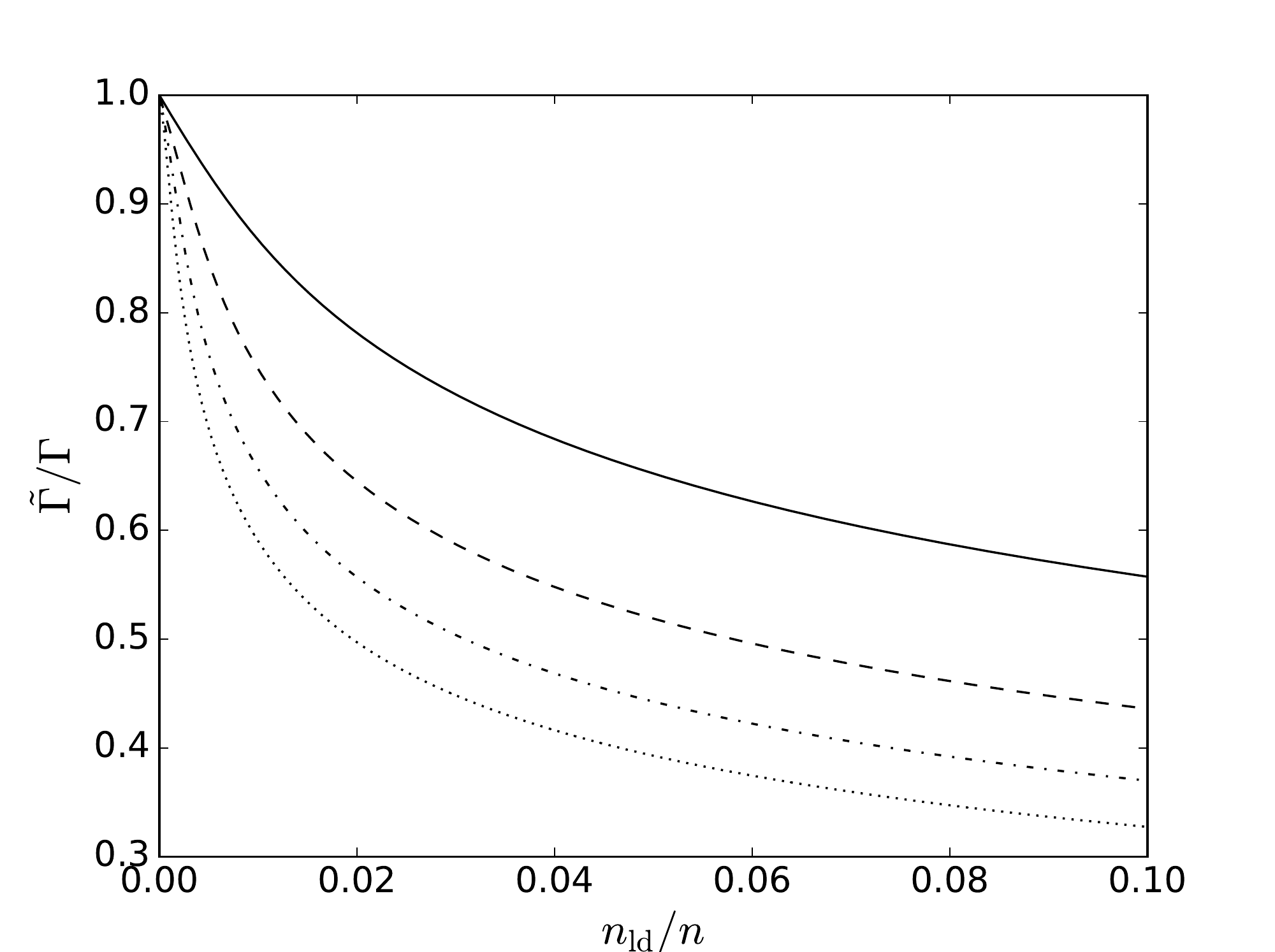} 
\caption{Relative Lorentz factor $\tilde{\Gamma}/\Gamma$ is plotted as a function of
relative loading particle number density $n_{\rm ld}/n$ for different initial $\Gamma$.
Solid curve corresponds to $\Gamma=5$, dashed to $\Gamma=10$, dash-dotted to $\Gamma=15$, and dotted to $\Gamma=20$.
} 
\label{fig_2}
\end{figure}

We see in Figure~\ref{fig_2} that for fairly moderate ratio $n_{\rm ld}/n$ the jet deceleration effect is appreciable. The effect is more
pronounced for the faster jets.

\subsection{Loaded flow magnetization}
\label{ss:LFM}

Now we are ready to discuss back reaction of the particle loading on the properties
of the magnetically dominated flow. We suppose that the magnetized jet consists of
electromagnetic field, cold particle flow 
(number density in the comoving reference frame $n$, relativistic enthalpy $\mu$) and
loading particles with the energy-momentum tensor $T_{\rm ld}^{ik}$  (\ref{Tik}), both
components having the same hydrodynamical velocity $\langle{\bf v}\rangle$.

General expressions for the energy $E(\Psi)$ and angular momentum
$L(\Psi)$ flux of the ideal relativistic MHD flow with anisotropic pressure \citep{AB-83, Ts-95, Kuz-05} include standard
anisotropic pressure parameter
\begin{equation}
\beta_{\rm a} = 4 \pi \frac{P_{n}-P_{s}}{b^2}.
\label{beta}
\end{equation}
In our case $\beta_{\rm a} < 0$. For steady-state flow the energy conservation is written as
\begin{equation}
{\rm div}\;{\bf S}=0,
\end{equation}
where $S^{\,i}=T^{0\,i}$. For the case of a cylindrical axisymmetric flow for the 
given~\citet{Lich-67} vector (see Appendix~B) with $b^0=0$ it transforms into 
\begin{equation}
\frac{\partial}{\partial z}\left(\frac{{\bf b}^2}{4\pi}+P_{\rm s}+\varepsilon_{\rm ld}+\varepsilon_{\rm bulk}\right)\tilde{\Gamma}^2\tilde{v}_{\rm z}=0,\label{divS}
\end{equation}
and after integration over $z$ we obtain
\begin{equation}
E(\Psi)B_{\rm p}={\rm const}
\end{equation}
with energy flux
\begin{equation}
E(\Psi)=\frac{\tilde{\Omega}_{\rm F}\tilde{I}}{2 \pi c}(1+|\beta_{\rm a}|)
+ \mu_{\rm ld}\eta_{\rm ld}\tilde{\Gamma} + \mu\eta\tilde{\Gamma}.
\label{EPsi}
\end{equation}
Each of four terms in (\ref{EPsi}) originates from the appropriate term in (\ref{divS}).
The angular momentum flux is
\begin{equation}
L(\Psi)=\frac{\tilde{I}}{2 \pi}(1+|\beta_{\rm a}|)
+ \mu_{\rm ld}\eta_{\rm ld} r_{\perp} \tilde{u}_{\varphi} + \mu\eta r_{\perp} \tilde{u}_{\varphi}.
\label{LPsi}
\end{equation}

Expressions (\ref{EPsi}) and (\ref{LPsi}) are sums of standard relations for cold flow 
(\ref{defEL}) and appropriate terms for loaded charged pairs. We must note here that angular velocity $\tilde{\Omega}_{\rm F}$
and current density $\tilde{I}$ are defined by the screened electric and magnetic fields. 
Here particle-to-magnetic flux ratios for initial $\eta(\Psi)$ and loading
$\eta_{\rm ld}(\Psi)$ plasma are determined by the standard relations
\begin{equation}
\begin{array}{l}
\displaystyle{\bf u} = \frac{\eta}{n}\tilde{\bf B} + \tilde{\Omega}_{\rm F} r_{\perp}\tilde{\Gamma}{\bf e}_{\varphi},\\ \ \\
\displaystyle{\bf u}_{\rm ld} = \frac{\eta_{\rm ld}}{n_{\rm ld}}\tilde{\bf B} + \tilde{\Omega}_{\rm F} r_{\perp} \tilde{\Gamma} {\bf e}_{\varphi}.
\end{array}
\label{defu}
\end{equation}
Since $\langle{\bf u}_{\rm ld}\rangle={\bf u}$, one can conclude that 
$\eta_{\rm ld}=\eta n_{\rm ld}/n$. Finally, $\mu = \varepsilon/n = m c^2$ and
$\mu_{\rm ld} = \varepsilon_{\rm ld}/n_{\rm ld} = m c^2 \tilde{\Gamma}$. 
In (\ref{EPsi})--(\ref{LPsi}) the first terms $\tilde{\Omega}_{\rm F}\tilde{I}/2\pi c$ and $\tilde{I}/2\pi$
correspond to the electromagnetic flux, while the second and the third ones describe
the fluxes corresponding to anisotropic pressure and internal energy of the loading
particles respectively. The last terms correspond to cold wind.

Taking into account the results of Subsection~\ref{ss:MLFP}, we can find a loaded flow magnetization. 
As secondary particles \nv{in the simple problem stated in subsection~\ref{ss:PM} carry zero energy flux (\ref{EPsi}), we may equate the corresponding integrals $E$ before and after loading}:
\begin{equation}
\frac{\Omega_{\rm F}I}{2 \pi c}+\mu\eta\Gamma=\frac{\tilde{\Omega}_{\rm F}\tilde{I}}{2 \pi c}(1+|\beta_{\rm a}|)
+ \mu_{\rm ld}\eta_{\rm ld}\tilde{\Gamma} + \mu\eta\tilde{\Gamma}.\label{noDeltaE}
\end{equation}
This expression allows us to calculate the magnetization of the mass loaded flow.
Indeed, one can rewrite the Poynting flux using standart functions $\tilde\Omega_{\rm F}$ and $\tilde I$ as
\begin{equation}
\tilde{\bf S}=\frac{1}{2\pi c}\tilde{\Omega_{\rm F}}\tilde{I}\tilde{\bf B}_{\rm p}.
\end{equation}
This corresponds to the first term in the r.h.s. in energy flux (\ref{noDeltaE}). The other three terms are 
anisotropic pressure of loaded flow, motion of loaded relativistic plasma, and bulk motion of jet plasma.
We must stress that the term with parameter $|\beta_{\rm a}|$, although being traditionally written through 
the Poynting flux, is related to the plasma internal energy, carried by loaded particles (see equation~(\ref{divS})).
Thus, the definition of magnetization parameter after loading is
\begin{equation}
\tilde{\sigma}=\left(\frac{\tilde{\Omega}_{\rm F}\tilde{I}}{2\pi c}\right)
\left(\frac{\tilde{\Omega}_{\rm F}\tilde{I}}{2\pi c}|\beta_{\rm a}|+\mu_{\rm ld}\eta_{\rm ld}\tilde\Gamma+\mu\eta\tilde\Gamma\right)^{-1}.
\label{def-tilde-sigma}
\end{equation} 
Using (\ref{noDeltaE}), we finally obtain the relation for the magnetization of loaded flow as
\begin{equation}
\tilde\sigma = {\sigma}\left(\frac{\tilde S}{S}\right)
\left[\frac{\tilde\Gamma}{\Gamma}\left(1+\tilde\Gamma\frac{n_{\rm ld}}{n}\right)+
\frac{n_{\rm ld}}{n}\frac{\tilde\Gamma^3}{\Gamma^2}\frac{\tilde\beta^3}{2\beta}\right]^{-1}.
\end{equation}

\begin{figure}
\includegraphics[scale=0.4]{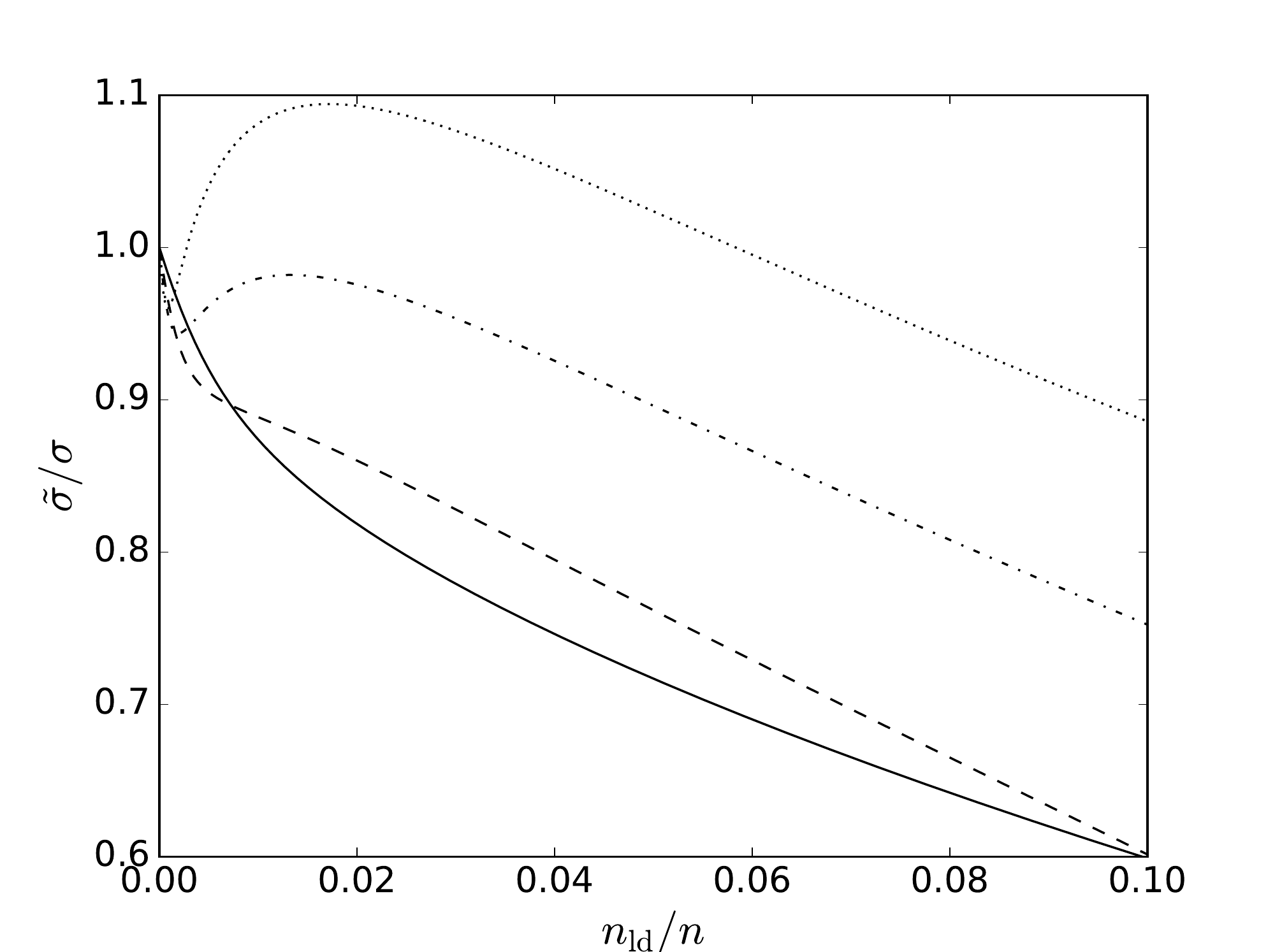} 
\caption{Relative magnetization $\tilde{\sigma}/\sigma$ is plotted as a function of
relative loading particle number density $n_{\rm ld}/n$ for different initial $\Gamma$.
Solid curve corresponds to $\Gamma=5$, dashed to $\Gamma=10$, dash-dotted to $\Gamma=15$, and dotted to $\Gamma=20$.
} 
\label{fig_3}
\end{figure}

We plot in Figure~\ref{fig_3} the loaded magnetization $\tilde\sigma$ as a function of loading 
particles number density $n_{\rm ld}$ for different initial bulk flow Lorentz factors $\Gamma$. 
As one can see, for $\Gamma=5$ and $10$ the magnetization drops. However, for greater Lorentz 
factors it grows and becomes more than the unity. It has been noted in Subsection~\ref{ss:MLFP} 
that the charge loading screen the electric field greater than the manetic field. As a consequence, 
the Poynting flux of a flow always drops. But the initially faster flow decelerates more effectively, 
so for greater Lorentz factors the drop in kinetic energy flux is greater than that in Poynting 
flux, which leads to the growth of the flow magnetization. For $n_{\rm ld}/n\rightarrow +0$, the 
magnetization always decreases as
\begin{equation}
\frac{\tilde\sigma}{\sigma}=1-\left(\frac{3}{2}+\frac{3}{\sigma}\right)\frac{n_{\rm ld}}{n}\Gamma\approx 1-\frac{9}{2}\frac{n_{\rm ld}}{n}\Gamma,
\end{equation}
which can be seen in Figure~\ref{fig_3}. This implies that for small enough loading particle 
number density the fields screening, and so the diminishing of Poynting flux, is more important 
than the flow deceleration itself.

We must stress again that the screening of electric and magnetic fields is important for arbitrarily low 
pair loading. Indeed, suppose that $q\rightarrow +0$, and so $\tilde{\Gamma}\rightarrow\Gamma-0$. In this case
we can retain in equation (\ref{eq_for_gamma}) the leading terms
\begin{equation}
16\tilde{\Gamma}^6\Gamma^4-16\tilde{\Gamma}^{10}-48q\tilde{\Gamma}^9\Gamma^2=0
\end{equation}
and obtain
\begin{equation}
\tilde{\Gamma}=\Gamma-\frac{3}{\sigma}\frac{n_{\rm ld}}{n}\Gamma^2. \label{linear_case}
\end{equation}
This relation corresponds to a tangent line to any curve in Figure~\ref{fig_2} at $n_{\rm ld}/n=0$.

On the other hand, we may propose that for extremely low loaded particles number density
the effects of the electric and magnetic fields screening is negligible, and the angular velocity $\Omega_{\rm F}$ and
current $I$ do not change. Under this assumption
one can obtain from (\ref{noDeltaE})
\begin{equation}
\Gamma-\tilde{\Gamma}=\frac{3}{2\sigma}\frac{n_{\rm ld}}{n}\Gamma^2,
\end{equation}
which is in contradiction with (\ref{linear_case}). This implies that the effect of
screening of the electric and magnetic fields is important for any rate of
pairs loading, and, apparently, it has not been included into consideration by the previous works.
When the loading particles number density tends to zero, the deceleration rate is greatest and can be expressed in the following
short form
\begin{equation}
\frac{\dot{\Gamma}}{\Gamma}=-\frac{3\Gamma}{2\sigma}\frac{\dot{n}}{n}.
\end{equation}


\section{Effects of pairs created with arbitrary velocities}
\label{s:ArbVel}

In Section~\ref{s:PL} we have considered pairs created ar rest in the laboratory frame.
Let us consider now the effects of particles loading if pairs are created with their center of mass (c.m.) moving 
with the velocity $\beta_0$ and corresponding Lorentz factor $\Gamma_0$ along the jet. If the energies of 
interacting photons are greater than threshold, electron and positron are created with Lorentz factors $\gamma'_0$ 
and isotropically distributed velocities $\beta'_0$ in the c.m. frame. The direction of velocity for electron/positron
in c.m. frame is set by spherical angles $\theta'$ and $\varphi'$.
The initial conditions for such a pair are
\begin{equation}
\begin{array}{l}
\gamma_{0\pm}=\Gamma_0\gamma'_0\left(1\pm\beta_0\beta'_0\cos\theta'\right), \\ \ \\
u^{r}_{0\pm}=\mp\gamma'_0\beta'_0\sin\theta'\cos\varphi', \\ \ \\
u^{z}_{0\pm}=\Gamma_0\gamma'_0\left(\beta_0\pm\beta'_0\cos\theta'\right).
\end{array}\label{amic}
\end{equation}
Here plus and minus designate the positive and negative charge, as for them the initial conditions at the laboratory
frame are different.
The particle Lorentz factor and $r$-component of four-velocity are
\begin{equation}
\begin{array}{rcl}
\displaystyle\gamma_{\pm}(\tau)&=&\Gamma^2\left(\gamma_{0\pm}-\beta u_{0\pm}^{z}\right)+\\ \ \\
&+&\displaystyle\cos\Omega'\tau\Gamma^2\beta\left(u_{0\pm}^z-\beta\gamma_{0\pm}\right)\pm\\ \ \\
\displaystyle&\pm&\sin\Omega'\tau\beta\Gamma u_{0\pm}^r,\\ \ \\
u_{\pm}^r(\tau)&=&\cos\Omega'\tau u_{0\pm}^r\pm\sin\Omega'\tau\Gamma\left(u_{0\pm}^z-\beta\gamma_{0\pm}\right).
\end{array}\label{amv}
\end{equation} 
The $r$-coordinate for each charge is
\begin{equation}
\begin{array}{rcl}
r_{\pm}(\tau) & = & \displaystyle \left[R_0\pm\frac{c\Gamma}{\Omega'}\left(u^z_{0\pm}-\beta\gamma_{0\pm}\right)\right]\mp\\ \ \\
 & \mp & \displaystyle\cos\Omega'\tau\frac{c\Gamma}{\Omega'}\left(u^z_{0\pm}-\beta\gamma_{0\pm}\right)+\sin\Omega'\tau\frac{c}{\Omega'}u_{0\pm}^r.
\end{array}
\end{equation} 

In order to obtain the mean charge displacement, we need to average the particles displacement at every space point
over the particles ensemble. The direction of initial velocity of each charge in c.m. frame is isotropic. The trajectory
phase $\varphi=\Omega'\tau$ has a weight function $\gamma_{\pm}(\tau)$, since each charge spends different laboratory time
on each part of its gyration trajectory. Thus, the full distribution function is
\begin{equation}
f_{\pm}(\theta',\,\varphi',\,\varphi)=\gamma_{\pm}(\theta',\,\varphi',\,\varphi),
\end{equation}
with the normalization included explicitly.

Obviously, for a given pair the Lorentz factors of electron and positron are different, so we define
the average charge displacement as
\begin{equation}
\begin{array}{rcl}
\displaystyle r_{\perp}&=&\langle r_+\rangle-\langle r_-\rangle=\\ \ \\
\displaystyle&=&\left.\int r_+f_+d\Omega_{\theta'\varphi'}d\tau\right/\int f_+d\Omega_{\theta'\varphi'}d\tau-\\ \ \\
\displaystyle&-&\left.\int r_-f_-d\Omega_{\theta'\varphi'}d\tau\right/\int f_-d\Omega_{\theta'\varphi'}d\tau.
\end{array}
\end{equation}  
Using this expression, one can find the mean charge displacement
\begin{equation}
\begin{array}{rcl}
r_{\perp}&=&\displaystyle\frac{m_{\rm e}c^2\tilde{\Gamma}^2}{e\tilde{B}}\Gamma_0\gamma'_0\left[2(\beta_0-\tilde{\beta})\left(1+\frac{\beta_0^{'2}}{3}\right)-\right.\\ \ \\
&-&\displaystyle \left.\tilde{\beta}\frac{(\beta_0-\tilde{\beta})^2}{1-\tilde{\beta}\beta_0}-\tilde{\beta}(1-\tilde{\beta}\beta_0)\frac{\beta_0^{'2}}{3}\right].\label{r_perp_arb}
\end{array}
\end{equation}
Three different limits can be regarded for the problem.

\noindent
{\it 1. Deceleration along the jet due to fields screening}
\label{ss:Dec}

The first case corresponds to condition $\Gamma_0 \ll \Gamma$. As the limiting case we set $\Gamma_0=1$ which leads to
\begin{equation}
r_{\perp}=-\frac{m_{\rm e}c^2\tilde{\beta}\tilde{\Gamma}^2}{e\tilde{B}}\gamma'_0\left(2+\tilde{\beta}^2+\beta^{'2}_0\right).\label{rperp1}
\end{equation}
This is exactly the loading regarded in the previous section, but for the loaded particles with higher initial temperature. 
Thus, the effect of the fields screening is even more pronounced for
non-zero $\beta'_0$. Indeed, we see from equations~(\ref{rperp0}) and (\ref{rperp1}) that the average charge displacements $r_{\perp}$ 
differ by the factor
\begin{equation}
f_1=\frac{\gamma'_0(2+\tilde\beta^2+\beta_0^{'2})}{2+\tilde\beta^2}\ge 1.
\end{equation}
Thus, all the result of Subsection~\ref{ss:MLFP} are applicable if we divide the loading particle number density by
$f_1$. In particular, we need to load the flow by $f_1$-times less charge number density than in Subsection~\ref{ss:MLFP}
in order to achieve the same deceleration rate.
However, as the reaction cross-section behaves as $\sigma_{\gamma\gamma}\propto(\ln 2x-1)/x^2$, where $x$ is a photon
energy normalized on the electron rest-mass in c.m. system (see, e.g., \citet{Sven-82}), the pairs are more likely created with non-relativistic or mildly
relativistic speeds with respect to the c.m. system. 

We must note, that although the impact of charges creation described in Section~\ref{s:PL} and in Sect.~\ref{ss:Dec} with c.m. moving
slower than the jet initial bulk velocity are very illustrative, the physical conditions that lead to such a pair creation can hardly be 
imagined in the jet environment. So, we proceed to the cases that have astrophysical applications. 

\noindent
{\it 2. Pure mass-loading}
\label{ss:Pure}

The condition $\Gamma_0\approx\Gamma$ and $\beta_0\approx\beta$ corresponds to the case when center-of-mass of each pair 
is moving with the hydrodynamic velocity. As it can be seen from the general expression for the average charge displacement~(\ref{r_perp_arb}),
no charge separation and, consequently, no fields screening take place in this case. The effect of flow deceleration is purely due to mass loading. 

However, due to cylindrical jet geometry this will lead not to the flow acceleration, but to a flow deceleration.
Pairs created in a center-of-mass reference frame at rest with reference to bulk flow motion do not contribute
into electric and magnetic fields screening/enhancement. Thus, there is no change in integral
$\Omega_{\rm F}(\Psi)$, corresponding to the electric field, and in the total current $I$, corresponding to the magnetic field. 
The analysis of the effect of deceleration is even simpler in this case, and the deceleration rate can be obtained through the examination of
the expression for the integral $E(\Psi)$ solely.

Using the known charges motion (\ref{amic}) and (\ref{amv}) with $\Gamma_0=\Gamma$ and $\beta_0=\beta$, we
obtain
\begin{equation}
\begin{array}{rcl}
T_{\rm ld}^{00}&=&\displaystyle m_{\rm e}c^2n_{\rm ld}\Gamma^2\gamma'_0\left(1+\beta^2\frac{\beta^{'2}_0}{3}\right), \\ \ \\
T_{\rm ld}^{11}&=&\displaystyle m_{\rm e}c^2n_{\rm ld}\gamma'_0\frac{\beta^{'2}_0}{3}.
\end{array}
\end{equation}
From these we get the internal energy density and pressure
\begin{equation}
\begin{array}{rcl}
\varepsilon_{\rm ld}&=&\displaystyle m_{\rm e}c^2n_{\rm ld}\gamma'_0, \\ \ \\
P_{\rm s}&=&\displaystyle m_{\rm e}c^2n_{\rm ld}\gamma'_0\frac{\beta^{'2}_0}{3}.
\end{array}\label{Ps_puremass}
\end{equation}

\begin{figure}
\includegraphics[scale=0.4]{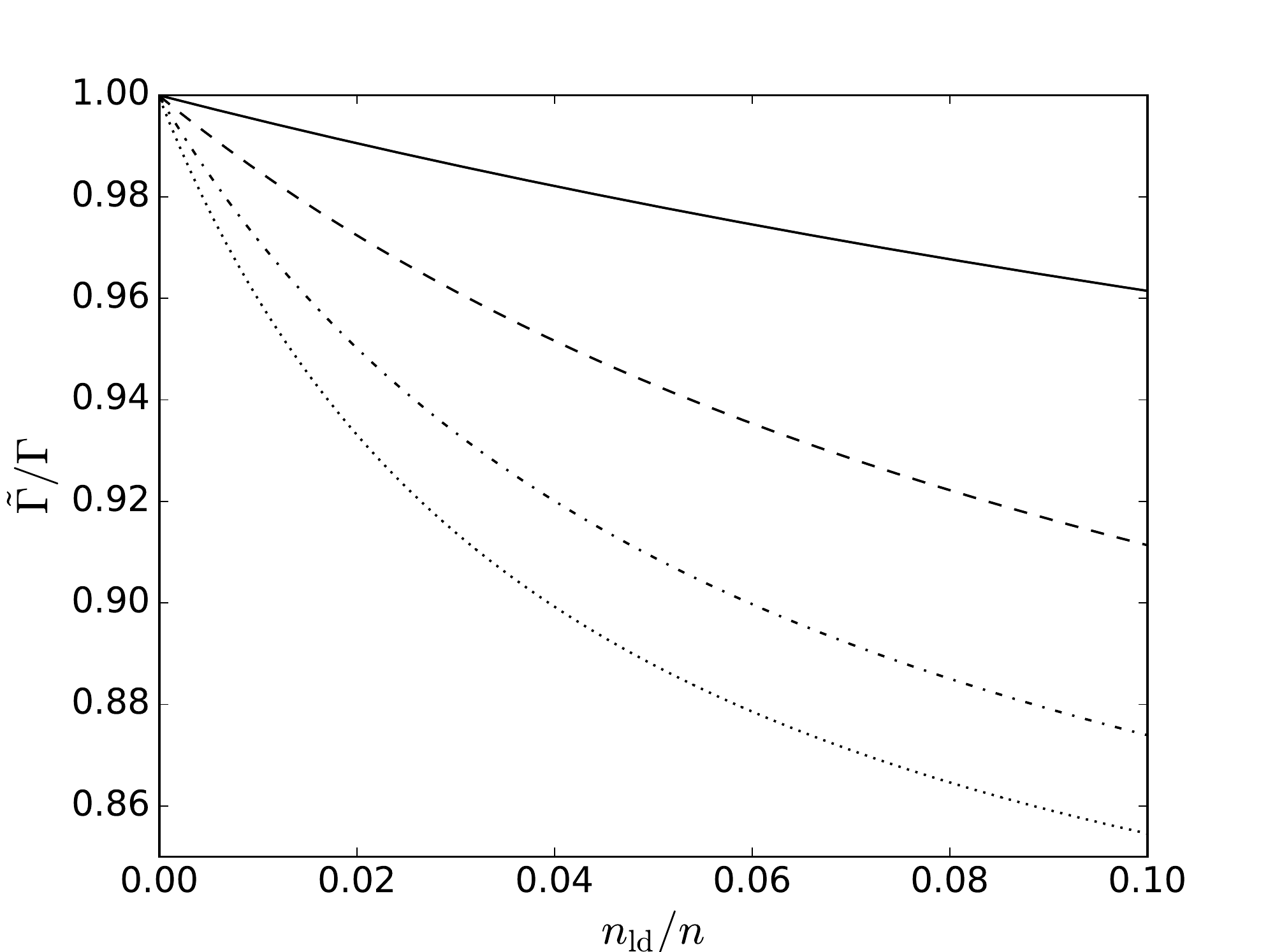} 
\caption{The deceleration due to pure mass-loading. Ratio of the Lorentz factors after and before loading is plotted as a function of the relative 
loading particle number density $n_{\rm ld}/n$ for different $\gamma'_0$.
Solid curve corresponds to $\gamma'_0=2$, dashed to $\gamma'_0=5$, dash-dotted to $\gamma'_0=10$, and dotted to $\gamma'_0=2$.
} 
\label{fig_4}
\end{figure}

\begin{figure}
\includegraphics[scale=0.4]{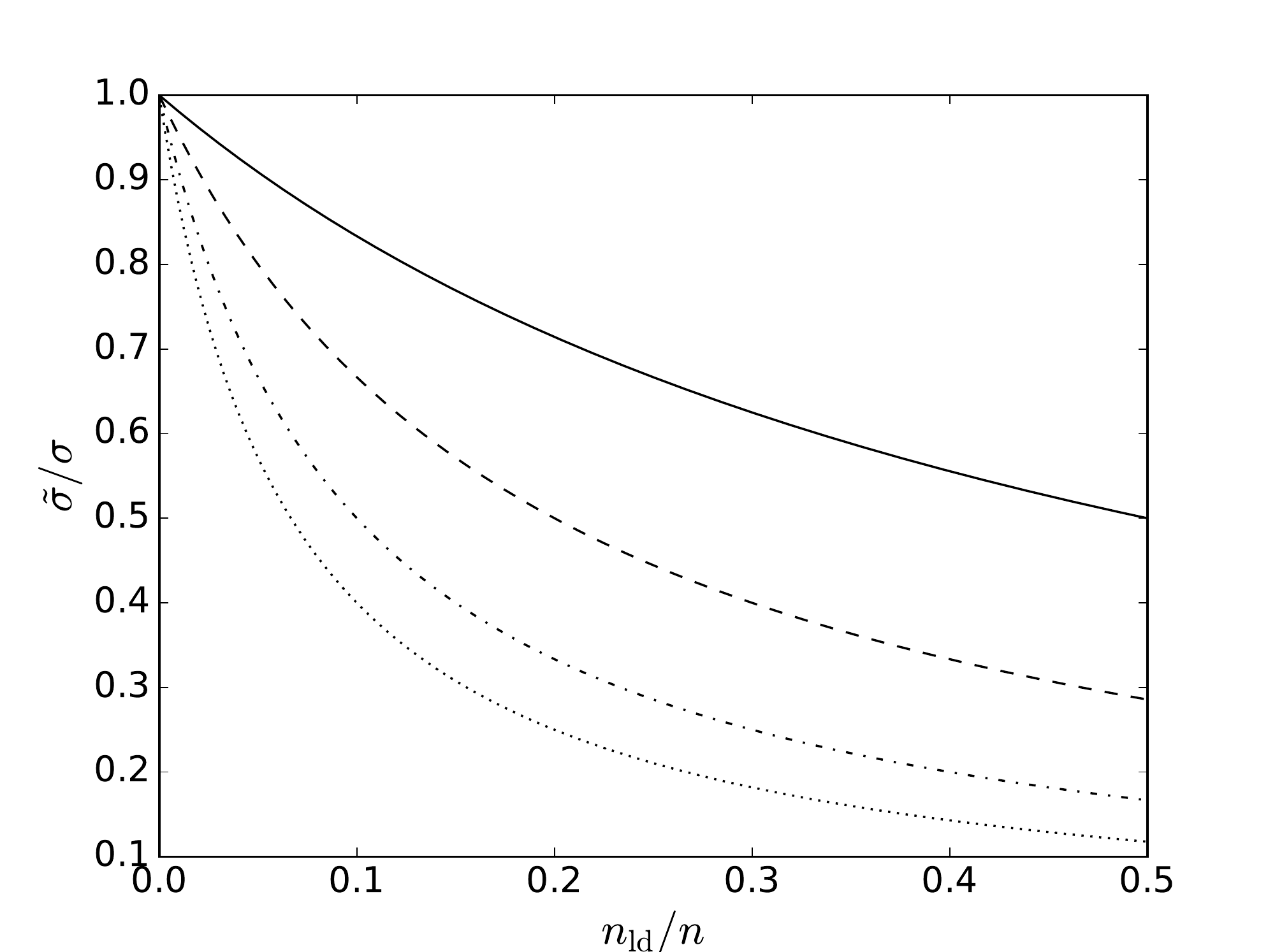} 
\caption{The deceleration due to pure mass-loading. Ratio of the flow magnetization after and before loading is plotted as a function of the relative 
loading particle number density $n_{\rm ld}/n$ for different $\gamma'_0$.
Solid curve corresponds to $\gamma'_0=2$, dashed to $\gamma'_0=5$, dash-dotted to $\gamma'_0=10$, and dotted to $\gamma'_0=2$.
} 
\label{fig_5}
\end{figure}

The particle energy flux before mass loading is defined by Eqn. (\ref{defEL}), and after mass loading by Eqn. (\ref{EPsi}). But
the total change in the energy flux is equal to energy flux of created pairs
\begin{equation}
\Delta E=m_{\rm e}c^2\eta\gamma'_0\frac{n_{\rm ld}}{n}\Gamma.
\end{equation}
Thus, we \nv{equate the energy fluxes before and after loading:} 
\begin{equation}
\begin{array}{l}
\displaystyle\frac{\Omega_{\rm F}I}{2\pi c}+\displaystyle\mu\eta\Gamma+\Delta E=\\ \ \\
=\displaystyle\frac{\Omega_{\rm F}I}{2\pi c}
\left(1+\frac{4\pi\tilde\Gamma^2}{B^2}P_{\rm s}\right)+\mu\eta\tilde\Gamma+\mu\eta\gamma'_0\frac{n_{\rm ld}}{n}\tilde\Gamma.
\end{array}
\end{equation}
Obviously, for $\beta'_0=0$ there is no change in the flow Lorentz factor because the cold plasma loaded with the average drift velocity 
does not contribute to the change in the bulk flow velocity. For relativistic plasma
we obtain the following relation on $\tilde\Gamma$ of decelerated flow for given $\beta'_0$ and $n_{\rm ld}$:
\begin{equation}
\frac{\tilde\Gamma^2}{\Gamma^2}\frac{n_{\rm ld}}{n}\gamma'_0\frac{\beta^{'2}_0}{3}+
\frac{\tilde\Gamma}{\Gamma}\left(1+\gamma'_0\frac{n_{\rm ld}}{n}\right)=1+\gamma'_0\frac{n_{\rm ld}}{n}.\label{puremass-1}
\end{equation}
The form of this equation suggests that the rate of the flow deceleration due to pure mass loading does not
depend on $\Gamma$, but on loading particles temperature, which is characterized by $\gamma'_0$.
The result is shown in Fig.~\ref{fig_4}. For $n_{\rm ld}/n\ll 1$ the deceleration rate is given by
\begin{equation}
\frac{\dot\Gamma}{\Gamma}=-\frac{\dot{n}}{n}\gamma'_0\frac{\beta^{'2}_0}{3}.\label{dot-n-pure}
\end{equation}
Thus, the flow magnetization due to pure mass loading is defined by
\begin{equation}
\frac{\tilde\sigma}{\sigma}=\left[\frac{\tilde\Gamma}{\Gamma}+\frac{n_{\rm ld}}{n}\gamma'_0\left(\frac{\tilde\Gamma}{\Gamma}+
\frac{\tilde\Gamma^2}{\Gamma^2}\frac{\beta^{'2}_0}{3}\right)\right]^{-1}
\end{equation}  
and it is always less than unity (see Fig.~\ref{fig_5}).

\noindent
{\it 3. Acceleration along the jet due to fields enhancement}
\label{ss:Acc}

If the center-of-mass velocity of a pair is much greater than the jet bulk velocity $\Gamma_0\gg\Gamma$, we obtain using (\ref{r_perp_arb}) the following expression for the average charge displacement
\begin{equation}
r_{\perp}=\frac{m_{\rm e}c^2}{e\tilde{B}}\frac{\Gamma_0\gamma'_0}{2}\left(1+\frac{\beta^{'2}_{0}}{3}\right).\label{rperp2}
\end{equation}
In this case average charge separation is positive. This implies
that instead of electric and magnetic fields screening the enhancement of the fields takes place. The flow must accelerate due to
such charge movement. This is easy to understand: opposite charges initially created in the same space point move in opposite direction 
depending on the initial velocity of center-of-mass in the jet proper frame where the Lorentz force is due to $B'$.
This induces a flow acceleration. 
 
\begin{figure}
\includegraphics[scale=0.4]{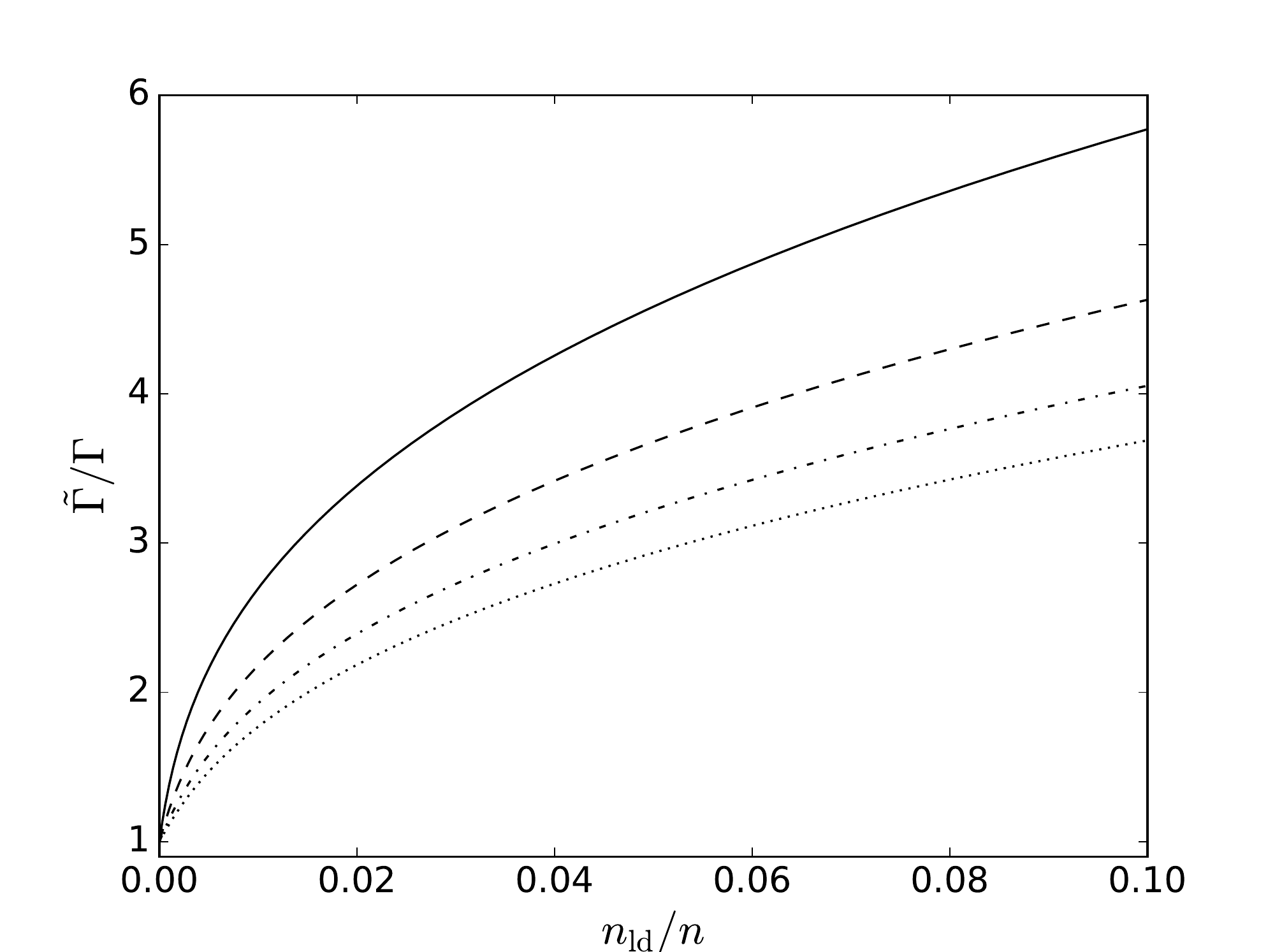} 
\caption{The plasma acceleration due to charge loading with pairs c.m. having $\Gamma_0=10^3$. 
Ratio of the Lorentz factors after and before loading is plotted as a function of the relative 
loading particle number density $n_{\rm ld}/n$ for different $\Gamma$.
Solid curve corresponds to $\Gamma=5$, dashed to $\Gamma=10$, dash-dotted to $\Gamma=15$, and dotted to $\Gamma=20$.
} 
\label{fig_6}
\end{figure}

The Lorentz factor of a flow after loading, calculated as corresponding to the drift velocity in disturbed fields, 
satisfies the following algebraic equation:
\begin{equation}
\begin{array}{l}
\displaystyle-16\tilde{\Gamma}^{6}+16\tilde{\Gamma}^4+
\tilde{\Gamma}^3 16q\beta\Gamma^2+\\ \ \\
\displaystyle+\tilde\Gamma^2 16\Gamma^2\left(\Gamma^2-1\right)+\tilde\Gamma\left(-8q\beta\Gamma^2\right)+q^2=0,
\end{array}\label{eq_for_gamma_acc}
\end{equation}
where 
\begin{equation}
q=\frac{4}{\sigma}\frac{n_{\rm ld}}{n}\Gamma_0\gamma'_0\left(1+\frac{\beta^{'2}_0}{3}\right).\label{q-acc}
\end{equation}
The appropriate solutions are presented in Fig.~\ref{fig_6}. As we see, the faster flow, the slower it accelerates.
For $n_{\rm ld}/n\ll 1$ the acceleration rate is
\begin{equation}
\frac{\dot\Gamma}{\Gamma}=\frac{1}{\Gamma\sigma}\frac{\dot{n}}{n}\Gamma_0\gamma'_0\left(1+\frac{\beta^{'2}_0}{3}\right).\label{n_dot_acc}
\end{equation} 

\begin{figure}
\includegraphics[scale=0.4]{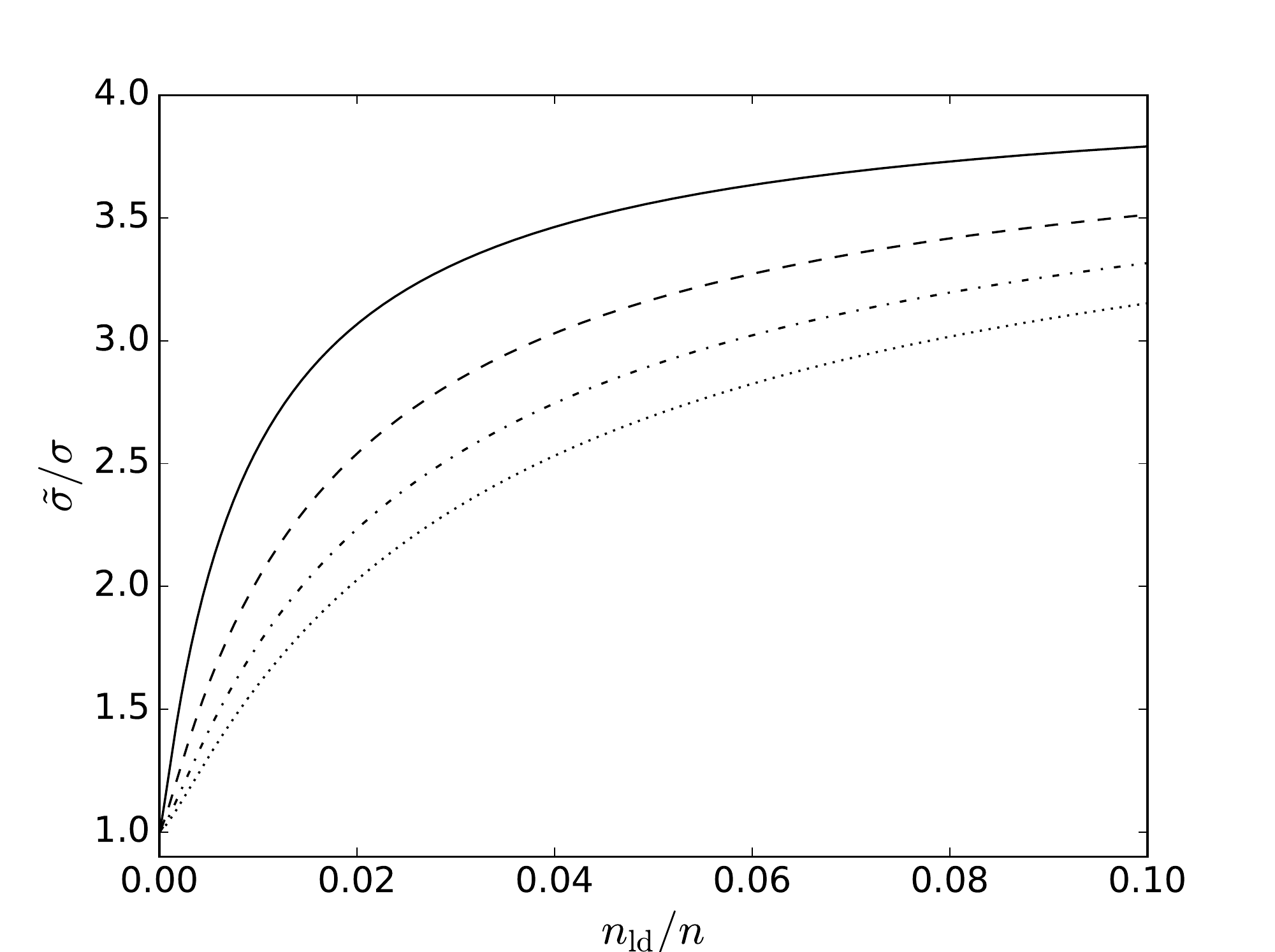} 
\caption{The plasma acceleration due to charge loading with pairs c.m. having $\Gamma_0=10^3$. 
Ratio of the flow magnetization after and before loading is plotted as a function of the relative 
loading particle number density $n_{\rm ld}/n$ for different $\Gamma$.
Solid curve corresponds to $\Gamma=5$, dashed to $\Gamma=10$, dash-dotted to $\Gamma=15$, and dotted to $\Gamma=20$.
} 
\label{fig_7}
\end{figure}

In order to compute the behavior of a flow magnetization, we calculate the internal energy and anisotropic pressure
through the corresponding energy-momentum tensor components. 
For $\Gamma_0\gg\Gamma$ one can obtain
\begin{equation}
\begin{array}{rcl}
\displaystyle T^{00}_{\rm ld}&=&\displaystyle m_{\rm e}c^2\Gamma n_{\rm ld}\left[\frac{1}{2}\Gamma_0\gamma'_0\left(1+\frac{\beta^{'2}_0}{3}\right)
\left(1+\frac{\beta^2}{2}\right)+\right.\\ \ \\
&+&\displaystyle\left.\Gamma^2\beta^2\frac{\gamma'_0}{\Gamma_0}\frac{\beta^{'2}_0}{3}\right],\\ \ \\
\displaystyle T^{11}_{\rm ld}&=&\displaystyle m_{\rm e}c^2\Gamma n_{\rm ld}\left[\frac{\gamma'_0}{\Gamma_0}\frac{\beta^{'2}_0}{3}+
\frac{\Gamma_0\gamma'_0}{4\Gamma^2}\left(1+\frac{\beta^{'2}_{0}}{3}\right)\right],
\end{array}\label{Ps_accel}
\end{equation}
and 
\begin{equation}
\begin{array}{rcl}
\varepsilon_{\rm ld}&=&\displaystyle m_{\rm e}c^2 n_{\rm ld}\frac{\Gamma_0\gamma'_0}{2\Gamma}\left(1+\frac{\beta^{'2}_{0}}{3}\right), \\ \ \\ 
P_{\rm s}&=&\displaystyle T^{11}_{\rm ld}.\label{acc-prop}
\end{array}
\end{equation}
Using now the magnetization definition (\ref{def-tilde-sigma}) and obtained flow parameters (\ref{acc-prop}), the following
expression for the ratio of magnetizations after and before loading can be written:
\begin{equation}
\frac{\tilde\sigma}{\sigma}=\frac{\tilde S\left/S\right.}{\tilde K\left/K\right.},
\end{equation}
where
\begin{equation}
\begin{array}{l}
\displaystyle\frac{\tilde K}{K}=\frac{\tilde\Gamma}{\Gamma}+\frac{n_{\rm ld}}{n}\left[\frac{\Gamma_0\gamma'_0}{2\tilde\Gamma}\left(1+\frac{\beta^{'2}_0}{3}\right)+\right.\\ \ \\
\displaystyle+\left.\frac{\tilde\beta}{\beta}\frac{\tilde\Gamma^3}{\Gamma^2}\left(\frac{\gamma^{'2}_0}{\Gamma_0}\frac{\beta^{'2}_0}{3}+
\frac{\Gamma_0\gamma'_0}{4\tilde\Gamma^2}\left(1+\frac{\beta^{'2}_0}{3}\right)\right)\right],
\end{array}
\end{equation}
and the ratio $\tilde S/S$ can be readily computed using the following expression for the enhanced fields
\begin{equation}
\tilde{B}=\frac{B}{2}\left[1+\sqrt{1+q\frac{\tilde{\Gamma}\tilde\beta}{\Gamma^2}}\;\right],
\label{tildeB1}
\end{equation}
\begin{equation}
\tilde{E}=\frac{E}{2}\left[1+\sqrt{1+q\frac{\tilde{\Gamma}\tilde\beta}{\Gamma^2-1}}\;\right],
\label{tildeE1}
\end{equation}
where $q$ is defined by (\ref{q-acc}). As is shown in Fig.~\ref{fig_7}, here again, although the plasma accelerates, the Poynting flux grows faster
after loading than the plasma kinetic energy, 
so the magnetization is greater than unity.

\nv{Here we should stress, that there are two different processes leading to a local plasma acceleration or deceleration.
The first one is connected with the average charge separation, and the second one --- with the non-zero loading plasma temperature for the 
pure mass-loading. In both cases the average velocity of each of created particles is equal to the fluid velocity. 
However, there is also a gyration motion which can be interpreted as the loading plasma temperature. This
effective temperature is defined by the initial Lorentz factor in the pair c.m. system $\gamma'_0$ and corresponding velocity $\beta'_0$.
The effects connected with the charge separation and the fields screening cannot be erased by cooling of loaded plasma due to 
radiation or instabilities as $\beta'_0\rightarrow 0$. Indeed, once the pair created, its average separation (\ref{rperp0}), (\ref{rperp1}) and (\ref{rperp2}) does not 
go to zero as $\beta'_0\rightarrow 0$. Thus, any process that wipes the gyration of loaded particles does not affect the fields screening and, consequently, 
the appropriate plasma acceleration or deceleration. It is not so with the pure mass loading: the effect of a flow deceleration for a pure mass loading
depends on the non-zero temperature (see (\ref{puremass-1}) and (\ref{dot-n-pure})), so the bulk plasma velocity does not change for $\beta'_0=0$.}

\section{Astrophysical application}
\label{s:AA}

The process that can provide the appreciable number of pairs is a conversion of high-energy jet radiation with photon energies $E_{\gamma}$, 
propagating along the jet within opening angle $1/\Gamma$ in the laboratory frame, on the external soft photon field with photon energies $\epsilon_{\rm soft}$.
The latter may be a reprocessed by environment thermal radiation of an accretion disk, as suggested by \citet{Sikora2}.
Such process would account for the jet acceleration, since the c.m. of these two photons have the Lorentz factor
\begin{equation}
\Gamma_0=\sqrt{\frac{E_{\gamma}}{4\epsilon_{\rm soft}}}.
\end{equation}
For soft photons in IR to UV range ($10^{-2}$ to $10^2$~eV) and reaction threshold condition, the Lorentz factor
is in $10^3\div 10^7$ range along the jet. Thus, the pairs are created with c.m. moving faster than cold plasma before the charge loading, 
accelerating the jet plasma, as it has been shown in Subsection~\ref{ss:Acc}. 
This process can operate naturally only on sub-parsec to parsec scales, since there we may expect strong enough soft photon field \citep{Sikora2}.

The probability of a pair creation due to two-photon conversion has been studied in number of works
by \citet{GSch-67, Sven-82, AAN-83, Zdz-88}.
The total pair creation rate due to reaction of $\gamma$-quanta with number density $N_{\gamma}$ on isotropically distributed
soft photons with number density $N_{\rm soft}$ is given by
\begin{equation}
\begin{array}{l}
\displaystyle\dot{n}_{\gamma\gamma}=\frac{c}{2}\sigma_{\rm T}N_{\gamma}N_{\rm soft}\int_0^{\infty}dx\,\tilde{n}_{\gamma}(x)\times\\ \ \\
\displaystyle\times\int_{1/x}^{\infty}dy\,\tilde{n}_{\rm soft}(y)\frac{1}{x^2y^2}\varphi(xy),\label{gg-1}
\end{array}
\end{equation}
where $\tilde{n}_{\gamma}(x)$ is a normalized to unity energy spectrum of $\gamma$-quanta depending on
dimensionless energy $x=E_{\gamma}/m_{\rm e}c^2$, $\tilde{n}_{\rm soft}(y)$ is the same
for soft photons with energy $\epsilon_{\rm soft}=ym_{\rm e}c^2$ \citep{Sven-82}.
Here
\begin{equation}
\varphi(xy)=\int_0^{xy}\frac{3}{8\pi r_0^2}\tilde\sigma_{\gamma\gamma}(s)s\,ds\label{gg-2}
\end{equation}
is averaged over solid angle $\gamma\gamma\rightarrow e^+e^-$ reaction cross-section,
$\sigma_{\rm T}=8\pi r_0^2/3$ is a Thomson cross-section, $r_0=e^2/m_{\rm e}c^2$ is a classical
electron radius, and $\sigma_{\rm T}\tilde\sigma_{\gamma\gamma}$ is a $\gamma\gamma\rightarrow e^+e^-$ cross-section
(see, e.g., \citet{RL-79, GSch-67, Sven-82, AAN-83, Zdz-88}). We use the asymptotic expression for $\varphi(xy)$, obtained by
\citet{GSch-67}. 
The reaction threshold condition is $xy\ge 1$. 

Number density of soft photons may be estimated as
\begin{equation}
N_{\rm soft}=\frac{L_{\rm soft}}{4\pi R_{\rm soft}^2c\langle\epsilon_{\rm soft}\rangle},
\end{equation} 
and for $\gamma$-quanta as
\begin{equation}
N_{\gamma}=\frac{L_{\gamma}}{S_{\gamma}c\langle E_{\gamma}\rangle}.
\end{equation}
Here $L_{\rm soft}$ and $L_{\gamma}$ are corresponding luminosities. The first one is distributed isotropically
over $4\pi R_{\rm soft}^2$, and for the second one we assume to be distributed over a jet cross-section in a laboratory frame.
The $\langle E \rangle$ are averaged over energy distribution of corresponding photon fields. 

Our aim is to obtain the needed to account for the observed jet accelerations and deceleration luminosities.
As the zero estimate we take the monoenergetic photon spectra. As $\varphi(xy)$ has a maximum of $0.21$ (in the units of Thomson 
cross-section) 
at $xy\approx 3.5$ \citep{Zdz-88}, we take as a rough model for the high energy conversion on soft photon field: 
$\epsilon_{\rm soft}=1.6\cdot 10^{-12}$~erg,
$E_{\gamma}=1.5$~erg. From these the pair c.m. Lorentz factor $\Gamma_0=4.8\cdot 10^5$, and the initial charge energy in the c.m. 
frame $\gamma'_0=1.8$ can be computed. This gives for the pairs production rate
\begin{equation}
\dot{n}_{\gamma\gamma}=8.8\cdot 10^{-10}\Gamma^2\frac{L_{\rm soft,\,45}L_{\gamma,\,45}}{R^2_{\rm soft,\,pc}R^2_{\gamma,\,\rm pc}}\;{\rm cm^{-3}\,s^{-1}}.\label{prod_rate1}
\end{equation}
Here $(L/{\rm erg\cdot s^{-1}})=10^{45}L_{45}$.

The reported by \citet{MOJAVE12} observed characteristic jet acceleration and deceleration rate 
is $\dot{\Gamma}/\Gamma\approx 10^{-3}$ to $10^{-2}$ per year 
in a galaxy frame, or 
\begin{equation}
\frac{\dot\Gamma}{\Gamma}=3.1\cdot 10^{-8}f\;{\rm s^{-1}},
\end{equation}
where $f$ is $10^{-3}$ to $10^{-2}$ is an observed factor of acceleration or deceleration.  
In order to account for such acceleration, we obtain from (\ref{n_dot_acc}) the corresponding pair creation rate
as
\begin{equation}
\dot{n}=2.9\cdot 10^{-13}fn\;{\rm cm^{-3}s^{-1}}.
\end{equation}
Thus, if we employ $n\approx 10^3$~${\rm cm^{-3}}$ on parsec distance \citep{AGN1}, 
$\Gamma\approx 10$ and $L_{\rm soft}\approx 10^{45}\;{\rm erg\,s^{-1}}$ \citep{SP06},
we need a luminosity of the order of $10^{40}\;{\rm erg\,s^{-1}}$ in TeV radiation of a jet
to explain the observed acceleration rate by the described above process.

On the other hand, the deceleration may also occur due to conversion of two photons of jet high energy radiation.
For this reaction the pairs are created with c.m. at rest in jet proper frame on average, since in the jet proper frame
the radiation can be regarded as isotropic. Thus, the results of Subsection~\ref{ss:Pure} are applicable, and a jet deceleration due to
pure mass-loading takes place. This process can operate on scales greater than a few parsec, where there is no substantial 
direct photon field from the disk or reprocessed by the clouds disk radiation.
If we want to account for the observed deceleration rates on scales of a few 10~pc, the pair creation threshold condition
implies that we need photons from the jet with energies greater than
$E_{\gamma}=1.5\cdot 10^{-6}$~erg. From expression (\ref{prod_rate1}), rewritten for the photons of a jet radiation, we obtain 
the pair production rate
\begin{equation}
\dot{n}_{\gamma\gamma}=8.8\cdot 10^{-12}L^2_{\gamma,\,45}\;{\rm cm^{-3}\,s^{-1}}.
\end{equation}  
Using (\ref{dot-n-pure}) for the factor $f$, reported by \citet{MOJAVE12}, we get $\dot{n}/n=7.8\cdot(10^{-12}\div 10^{-14})\;{\rm s^{-1}}$.
Thus, to account for the observed deceleration rates,
we need a luminosity in the laboratory frame $L\approx(3\div 9)10^{44}\;{\rm erg\,s^{-1}}$ at
MeV photons at a distance of 10~pc .

\section{Results and discussions}
\label{sect:diss}

The analysis of kinematic properties of bright features of relativistic jets~\citep{MOJAVE12} showed
that the systematic accelerations and decelerations of bright features is normal among the jets.
We propose a mechanism related to radiation that may change locally but significantly the bulk flow
velocity thus leading to the observed accelerations and decelerations. 

The two-photon pair conversion process supplies a jet with charges of number density that can be estimated using the
standard relations (\ref{gg-1}) and (\ref{gg-2}) \citep{GSch-67, Sven-82, Zdz-88}, i.e., the process of
mass and charge loading. We show that the latter is more important, since
it leads to magnetic and electric fields screening or enhancement, with the consequent influence on the 
flow motion in the domain of pair creation. We have showed that important role in the process is connected with the specific 
charges motion in the crossed electric and magnetic field 
and with fields screening or enhancement. The charge loading leads to a local change in such
MHD outflow parameters as the angular velocity $\Omega_{\rm F}(\Psi)$ and the total current $I$.
\nv{It is necessary to stress that such an initial charge separation is not wiped out with the cease of gyro-motion due to radiation or instabilities,
thus, the effect of fields screening is unaffected by it.}

The proposed mechanism works as a kind of radiation friction. Indeed, if the radiation field
is such as pair center-of-mass reference frame moves faster than the bulk flow, the charges 
gyration leads to the electric and magnetic fields enhancement and the flow acceleration. For the 
pairs created due to hard gamma-quanta of a jet radiation with the external soft isotropic radiation field,
the center-of-mass Lorentz factor exceeds $10^3$, which leads to extremely effective local flow acceleration
even for loading pairs number density as low as tenths of a per cent of initial particles number density.
Such pairs are mainly created close enough to the jet origin where there external radiation from 
the accretion disk provides soft external radiation field \citep{Sikora2}.    
For the pairs with center-of-mass moving slower than the bulk flow velocity,
local deceleration takes place either due to partial screening of electric and magnetic fields (see Subsection~\ref{ss:MLFP}), 
or due to pure mass loading without significant fields screening (see Subsection~\ref{ss:Pure}). The latter case
is more probable as in the far jet domains there is no significant external radiation field, and
pairs are created mainly due to interaction of internal jet photons, with center-of-mass moving with the
jet velocity on average. 

Thus, the effect may account for the observed by MOJAVE program statistically significant accelerations of 
bright features of a jets at distances less than 50 pc and decelerations at further distances from the jet core.

\nv{The suggested mechanism of charge loading may play a role of a source of instabilities in a jet.
The mirror instability \citep{RS-61} may wipe out the anisotropic pressure $P_{\rm s}$. However, as it has been stressed at the end of Section~\ref{s:ArbVel},
this will not affect the process of fields screening and associated with it fluid local acceleration or deceleration.
The mirror instability develops only for the weakly magnetized flows \citep{SK-93} --- when the ratio of plasma pressure
to magnetic pressure is greater than the unity. Let us apply this criteria to the loaded flow regarded in the previous sections. 
The anisotropic pressure $P_{\rm s}$ is exactly the $T^{11}_{\rm{ld}}$, with the index `1' corresponding to
the $r$--coordinate. Thus, it does not change with a transformation to the fluid frame. In the fluid frame the ratio of loaded plasma pressure to the
magnetic pressure can be written as
\begin{equation}
\beta_{\rm m}=\frac{P_{\rm s}}{B'^2/8\pi}.
\end{equation}
For the simplest case of a pair created at rest with respect to nucleus the expression for $P_{\rm s}$ (\ref{Ps}) gives
\begin{equation}
\beta_{\rm m}=\frac{1}{\sigma}\frac{n_{\rm ld}}{n}\Gamma,\label{bm1}
\end{equation}
the outflow stability condition $\beta_{\rm m}<1$ being fulfilled for the flow satisfying the condition $\delta B/B<1$ (\ref{DeltaBB}).
For the pure mass loading (\ref{Ps_puremass})
\begin{equation}
\beta_{\rm m}=\frac{2}{3}\frac{1}{\sigma}\frac{n_{\rm ld}}{n}\gamma'_0\beta_0^{'2},\label{bm2}
\end{equation}
and for pairs created with the center of mass moving faster than jet bulk velocity (\ref{Ps_accel})
\begin{equation}
\beta_{\rm m}=\frac{1}{\sigma}\frac{n_{\rm ld}}{n}\left[\frac{2}{3}\frac{\Gamma\gamma'_0\beta_0^{'2}}{\Gamma_0}+\frac{\Gamma_0\gamma'_0}{2\Gamma}
\left(1+\frac{\beta_0^{'2}}{3}\right)\right].\label{bm3}
\end{equation}
Equations (\ref{bm1}--\ref{bm3}) provide the criteria of the mirror stability of the loaded flow: it is stable
for $\beta_{\rm m}<1$.}

\nv{As the created charges' average velocity is equal to the fluid velocity, there is no current or 
beam of charged particles with respect to the flow in the pair creation region that can give rise to the Buneman \citep{Buneman-59, Iizuka-79} and 
two-stream instability.
However, after the region
of pair creation has been accelerated or decelerated, the presented above picture of a sharp boundary of the uniform pair
creation region may give rise to instabilities on the boundary loaded -- non-loaded flow. In this case the two-stream instability 
develops for the wave numbers $k<\omega_{\rm P}/v_0$, where $\omega_{\rm P}$ is a plasma frequency and $v_0$ is the velocity
of loaded plasma with respect to non-loaded plasma. The preliminary estimates give
\begin{equation}
v_0\approx\frac{c\Delta\Gamma}{\Gamma^3}.
\end{equation}   
Using the expression for $\Delta\Gamma$ (\ref{linear_case}) for the initial phase of the flow loading, we obtain
\begin{equation}
v_0\approx c\frac{3}{\sigma\Gamma}\frac{n_{\rm ld}}{n}.
\end{equation} 
So, unstable wave lengths are
\begin{equation}
\lambda>3\frac{c}{\omega_{\rm P}}\frac{n_{\rm ld}}{n}\frac{1}{\sqrt{\sigma\Gamma}}.
\end{equation}
The more realistic picture of gradual decrease of loaded pairs number density
towards the jet axis due to external photon absorption may probably eliminate this possible source of such instabilities. However,
this question deserves the separate analysis on the base of the proposed mass-loading description.}

\nv{These preliminary estimates show that the process of charge loading may lead to instabilities in the layer, where the mass and charge loading
occurs. The question of the possible impact of instabilities on the proposed process will be addressed in the future work.} 

On the other hand, the same mechanism may account for the sights of radiation. Indeed, due to external 
comptonized hard and external soft radiation and following pairs creation, the local velocity of a bulk flow
motion changes. This leads to possible appearance of internal shocks which in turn give rise to
particle non-thermal acceleration and following synchrotron radiation, the external radiation thus working as a trigger
to the internal radiation site.   


\section*{Acknowledgments}
We would like to acknowledge M.~Barkov, E.~Derishev, Ya.N.~Istomin and J.~Poutanen for useful comments.
\nv{We also thank the unknown referee for the remarks touching the stability question that helped to 
improve this paper.}
The study of a mass and charge loading and its effects on the bulk plasma motion is supported by 
Russian Fund for Basic Research (RFBR) grant 16-32-60074 mol\_a\_dk. 
The preliminary study of possible instabilities is supported by Russian Science Foundation grant 16-12-10051.

\appendix

\section{Averaging procedure}
\label{A1}

To average a $\tau$-dependent function $A(\tau)$ with respect to laboratory time one
has to use the following relation:
\begin{equation}
\langle A\rangle=\frac{1}{T}\int_0^T A(\tau){\rm d}t=\frac{1}{\int_0^{T'}\gamma(\tau){\rm d}\tau}\int_0^{T'}A(\tau)\gamma(\tau){\rm d}\tau.
\end{equation}
Here $T'$ is an appropriate period in the loading particle rest frame.

The avaraging procedure for components of the energy-momentum tensor $T^{ik}$ is as follows.
$T^{ik}_{\rm ld} = mc^2\langle n^{\rm rest}_{\rm ld} u^{i}u^{k}\rangle$ contains the thermodynamical parameters in the particle rest frame.
For instance, the appropriate particle number density $n^{\rm rest}_{\rm ld}$ is related to the particle number density $n^{\rm lab}_{\rm ld}$ in
the laboratory frame by $n^{\rm lab}_{\rm ld}=\gamma(\tau)n^{\rm rest}_{\rm ld}$. The components $T^{ik}_{\rm ld}$ after averaging procedure has
to be written using the thermodynamical parameters in the jet reference frame, i.e. $n_{\rm ld}$ --- particle number density in the jet frame.
As $n^{\rm lab}_{\rm ld}=\Gamma n_{\rm ld}$, we write the energy-momentum tensor components as
\begin{equation}
T^{ik}_{\rm ld}=mc^2\langle n^{\rm rest}_{\rm ld} u^{i}u^{k}\rangle=mc^2\Gamma n_{\rm ld}\left\langle\frac{u^i u^k}{\gamma(\tau)}\right\rangle.
\end{equation}
Finally,
\begin{equation}
T^{ik}_{\rm ld}=mc^2\Gamma n_{\rm ld}\frac{1}{\int_0^{T'}\gamma(\tau){\rm d}\tau}\int_0^{T'}u^i(\tau)u^k(\tau){\rm d}\tau.
\end{equation}

\section{Energy-momentum tensor for loaded particles}
\label{A2}

The non-zero components of the energy momentum tensor $T^{ik}_{\rm ld}=\langle mc^2n^{\rm rest}_{\rm ld} u^{i}u^{k}\rangle$ 
in the laboratory reference frame obtained by the averaging procedure
are following:
\begin{eqnarray}
T^{00}_{\rm ld}&=&mc^2\Gamma^3n_{\rm ld}\left(1+\frac{\beta^4}{2}\right), \\
T^{02}_{\rm ld}&=&mc^2\Gamma^3n_{\rm ld}\beta\sin\alpha\left(1+\frac{\beta^2}{2}\right), \\
T^{03}_{\rm ld}&=&mc^2\Gamma^3n_{\rm ld}\beta\cos\alpha\left(1+\frac{\beta^2}{2}\right), \\
T^{11}_{\rm ld}&=&mc^2\Gamma n_{\rm ld}\frac{\beta^2}{2}, \\
T^{22}_{\rm ld}&=&mc^2\Gamma^3n_{\rm ld}\sin^2\alpha\frac{3\beta^2}{2}, \\
T^{23}_{\rm ld}&=&mc^2\Gamma^3n_{\rm ld}\sin\alpha\cos\alpha\frac{3\beta^2}{2}, \\
T^{33}_{\rm ld}&=&mc^2\Gamma^3n_{\rm ld}\cos^2\alpha\frac{3\beta^2}{2}.
\label{Tik-calc}
\end{eqnarray}

The components of hydrodynamical four-velocities in laboratory frame are
\begin{eqnarray}
U^0&=&\Gamma, \\
U^1&=&0, \\
U^2&=&\beta\Gamma\sin\alpha, \\
U^3&=&\beta\Gamma\cos\alpha.
\end{eqnarray}

Lichnerowicz vector components are

\begin{equation}
b^i=\left\{0;\;0;\;-\frac{B}{\Gamma}\sin\alpha;\;\frac{B}{\Gamma}\cos\alpha\right\}.
\label{hi-calc}
\end{equation}

\section{Algebraic equation on selfconsistent Lorentz factor of charge loaded outflow}
\label{A3}

When the screening/enhancement of the fields takes place it is convenient to find the general algebraic equation
for the $\tilde{\Gamma}$. Let us introduce dimensionless factor $F$:
\begin{equation}
r_{\perp}=\frac{m_{\rm e}c^2}{e\tilde{B}}F.
\end{equation}
The exact expression for the self-similar charge loaded flow velocity is
\begin{equation}
\tilde{\beta}=\beta\frac{1+\sqrt{1+Q/(\Gamma^2-1)}}{1+\sqrt{1+Q/\Gamma^2}}.
\end{equation}
In a physically interesting case of $Q/\Gamma^2\ll 1$ general algebraic equation on 
$\tilde{\Gamma}$ for the given $r_{\perp}(F)$ is the following:
\begin{equation}
\begin{array}{rcl}
& & \displaystyle Q^3\tilde{\Gamma}^2\frac{\Gamma^6+7\Gamma^4-10\Gamma^2+4}{32\Gamma^4\left(\Gamma^2-1\right)^3}+\\ \ \\
&+& \displaystyle Q^2\tilde{\Gamma}^2\frac{-3\Gamma^2+2}{8\Gamma^2\left(\Gamma^2-1\right)^2}+\\ \ \\
&+& \displaystyle Q\tilde{\Gamma}^2\frac{1}{2\left(\Gamma^2-1\right)}-\tilde{\Gamma}^2+\Gamma^2=0,
\end{array}
\end{equation}
where parameter 
\begin{equation}
Q=\frac{4}{\sigma}\frac{n_{\rm ld}}{n}\tilde{\Gamma}F.
\end{equation}
This is a third order decomposition over $Q/\Gamma^2$ which reproduce with high accuracy the
exact equation (\ref{eq_for_gamma}) for the special case of pairs created at rest in a 
nucleus rest frame.

\end{document}